\documentstyle[12pt]{article}

\catcode`\@=11
\long\def\@makefntext#1{ 
\protect\noindent \hbox to 3.2pt {\hskip-.9pt
$^{{\ninerm\@thefnmark}}$\hfil}#1\hfill} 

\def\thefootnote{\fnsymbol{footnote}}
 \def\@makefnmark{\hbox to 0pt{$^{\@thefnmark}$\hss}}  

\def\ps@myheadings{\let\@mkboth\@gobbletwo
\def\@oddhead{\hbox{} 
\rightmark\hfil\ninerm\thepage}
\def\@oddfoot{}\def\@evenhead{\ninerm\thepage\hfil 
\leftmark\hbox{}}\def\@evenfoot{}
\def\sectionmark##1{}\def\subsectionmark##1{}}

\begin{document}

\newcommand{\symbolfootnote}{\renewcommand{\thefootnote}
	{\fnsymbol{footnote}}}
\renewcommand{\thefootnote}{\fnsymbol{footnote}}
\newcommand{\alphfootnote}
	{\setcounter{footnote}{0}
	 \renewcommand{\thefootnote}{\sevenrm\alph{footnote}}}

\newcounter{sectionc}\newcounter{subsectionc}\newcounter{subsubsectionc}
\renewcommand{\section}[1] {\vspace{0.6cm}\addtocounter{sectionc}{1}
\setcounter{subsectionc}{0}\setcounter{subsubsectionc}{0}\noindent
	{\bf\thesectionc. #1}\par\vspace{0.4cm}}
\renewcommand{\subsection}[1] {\vspace{0.6cm}\addtocounter{subsectionc}{1}
	\setcounter{subsubsectionc}{0}\noindent
	{\it\thesectionc.\thesubsectionc. #1}\par\vspace{0.4cm}}
\renewcommand{\subsubsection}[1]
{\vspace{0.6cm}\addtocounter{subsubsectionc}{1}
	\noindent {\rm\thesectionc.\thesubsectionc.\thesubsubsectionc.
	#1}\par\vspace{0.4cm}}
\newcommand{\nonumsection}[1] {\vspace{0.6cm}\noindent{\bf #1}
	\par\vspace{0.4cm}}

\newcounter{appendixc}
\newcounter{subappendixc}[appendixc]
\newcounter{subsubappendixc}[subappendixc]
\renewcommand{\thesubappendixc}{\Alph{appendixc}.\arabic{subappendixc}}
\renewcommand{\thesubsubappendixc}
	{\Alph{appendixc}.\arabic{subappendixc}.\arabic{subsubappendixc}}

\renewcommand{\appendix}[1] {\vspace{0.6cm}
        \refstepcounter{appendixc}
        \setcounter{figure}{0}
        \setcounter{table}{0}
        \setcounter{equation}{0}
        \renewcommand{\thefigure}{\Alph{appendixc}.\arabic{figure}}
        \renewcommand{\thetable}{\Alph{appendixc}.\arabic{table}}
        \renewcommand{\theappendixc}{\Alph{appendixc}}
        \renewcommand{\theequation}{\Alph{appendixc}.\arabic{equation}}
        \noindent{\bf Appendix \theappendixc #1}\par\vspace{0.4cm}}
\newcommand{\subappendix}[1] {\vspace{0.6cm}
        \refstepcounter{subappendixc}
        \noindent{\bf Appendix \thesubappendixc. #1}\par\vspace{0.4cm}}
\newcommand{\subsubappendix}[1] {\vspace{0.6cm}
        \refstepcounter{subsubappendixc}
        \noindent{\it Appendix \thesubsubappendixc. #1}
	\par\vspace{0.4cm}}

\def\abstracts#1{{
	\centering{\begin{minipage}{30pc}\tenrm\baselineskip=12pt\noindent
	\centerline{\tenrm ABSTRACT}\vspace{0.3cm}
	\parindent=0pt #1
	\end{minipage} }\par}}

\newcommand{\bibit}{\it}
\newcommand{\bibbf}{\bf}
\renewenvironment{thebibliography}[1]
	{\begin{list}{\arabic{enumi}.}
	{\usecounter{enumi}\setlength{\parsep}{0pt}
\setlength{\leftmargin 1.25cm}{\rightmargin 0pt}
	 \setlength{\itemsep}{0pt} \settowidth
	{\labelwidth}{#1.}\sloppy}}{\end{list}}

\topsep=0in\parsep=0in\itemsep=0in
\parindent=1.5pc

\newcounter{itemlistc}
\newcounter{romanlistc}
\newcounter{alphlistc}
\newcounter{arabiclistc}
\newenvironment{itemlist}
    	{\setcounter{itemlistc}{0}
	 \begin{list}{$\bullet$}
	{\usecounter{itemlistc}
	 \setlength{\parsep}{0pt}
	 \setlength{\itemsep}{0pt}}}{\end{list}}

\newenvironment{romanlist}
	{\setcounter{romanlistc}{0}
	 \begin{list}{$($\roman{romanlistc}$)$}
	{\usecounter{romanlistc}
	 \setlength{\parsep}{0pt}
	 \setlength{\itemsep}{0pt}}}{\end{list}}

\newenvironment{alphlist}
	{\setcounter{alphlistc}{0}
	 \begin{list}{$($\alph{alphlistc}$)$}
	{\usecounter{alphlistc}
	 \setlength{\parsep}{0pt}
	 \setlength{\itemsep}{0pt}}}{\end{list}}

\newenvironment{arabiclist}
	{\setcounter{arabiclistc}{0}
	 \begin{list}{\arabic{arabiclistc}}
	{\usecounter{arabiclistc}
	 \setlength{\parsep}{0pt}
	 \setlength{\itemsep}{0pt}}}{\end{list}}

\newcommand{\fcaption}[1]{
        \refstepcounter{figure}
        \setbox\@tempboxa = \hbox{\tenrm Fig.~\thefigure. #1}
        \ifdim \wd\@tempboxa > 6in
           {\begin{center}
        \parbox{6in}{\tenrm\baselineskip=12pt Fig.~\thefigure. #1 }
            \end{center}}
        \else
             {\begin{center}
             {\tenrm Fig.~\thefigure. #1}
              \end{center}}
        \fi}

\newcommand{\tcaption}[1]{
        \refstepcounter{table}
        \setbox\@tempboxa = \hbox{\tenrm Table~\thetable. #1}
        \ifdim \wd\@tempboxa > 6in
           {\begin{center}
        \parbox{6in}{\tenrm\baselineskip=12pt Table~\thetable. #1 }
            \end{center}}
        \else
             {\begin{center}
             {\tenrm Table~\thetable. #1}
              \end{center}}
        \fi}

\def\@citex[#1]#2{\if@filesw\immediate\write\@auxout
	{\string\citation{#2}}\fi
\def\@citea{}\@cite{\@for\@citeb:=#2\do
	{\@citea\def\@citea{,}\@ifundefined
	{b@\@citeb}{{\bf ?}\@warning
	{Citation `\@citeb' on page \thepage \space undefined}}
	{\csname b@\@citeb\endcsname}}}{#1}}

\newif\if@cghi
\def\cite{\@cghitrue\@ifnextchar [{\@tempswatrue
	\@citex}{\@tempswafalse\@citex[]}}
\def\citelow{\@cghifalse\@ifnextchar [{\@tempswatrue
	\@citex}{\@tempswafalse\@citex[]}}
\def\@cite#1#2{{$\null^{#1}$\if@tempswa\typeout
	{IJCGA warning: optional citation argument
	ignored: `#2'} \fi}}
\newcommand{\citeup}{\cite}

\def\fnm#1{$^{\mbox{\scriptsize #1}}$}
\def\fnt#1#2{\footnotetext{\kern-.3em
	{$^{\mbox{\sevenrm #1}}$}{#2}}}

\font\twelvebf=cmbx10 scaled\magstep 1
\font\twelverm=cmr10 scaled\magstep 1
\font\twelveit=cmti10 scaled\magstep 1
\font\elevenbfit=cmbxti10 scaled\magstephalf
\font\elevenbf=cmbx10 scaled\magstephalf
\font\elevenrm=cmr10 scaled\magstephalf
\font\elevenit=cmti10 scaled\magstephalf
\font\bfit=cmbxti10
\font\tenbf=cmbx10
\font\tenrm=cmr10
\font\tenit=cmti10
\font\ninebf=cmbx9
\font\ninerm=cmr9
\font\nineit=cmti9
\font\eightbf=cmbx8
\font\eightrm=cmr8
\font\eightit=cmti8


\input epsf

\vspace{-.5in}
\hspace{4.2truein} UBCTP93-23
\vspace{.4in}

\centerline{\tenbf INTRODUCTION TO $Z_N$ SYMMETRY IN $SU_N$ GAUGE THEORIES}
\baselineskip=22pt
\centerline{\tenbf AT FINITE TEMPERATURES\footnote{Talk presented at the
3rd Thermal Fields Workshop held Aug 16 - 27, 1993 in
Banff, Canada.\\ This work was supported in part by the Natural Sciences and
Engineering Research Council of Canada.}}
\baselineskip=16pt
\vspace{0.8cm}
\centerline{\tenrm NATHAN WEISS}
\baselineskip=13pt
\centerline{\tenit Department of Physics, University of British Columbia}
\baselineskip=12pt
\centerline{\tenit Vancouver, British Columbia, V6T 1Z1, Canada}

\vspace{0.9cm}
\abstracts{There are several talks at this workshop on the physical
interpretation and  consequences of the $Z_N$ symmetry which
is present in the Euclidean Path Integral formulation of $SU_N$
Gauge Theories at finite temperature. The purpose of this paper
is to present an introduction to this subject. After a brief review
of Gluodynamics at finite temperature, the nature
of the $Z_N$ symmetry in this system is described and its relationship
to confinement is discussed.  $Z_N$ domain walls and bubbles
are then described as is their relationship to the confining--deconfining
phase transition. The effect of Fermions is then considered and
the presence of metastable extrema of the Effective Potential for
the Polyakov--Wilson Line is described. It is then argued that
these metastable extrema do not correspond to physically
realizable metastable states. }

\vfil
\twelverm   
\baselineskip=14pt
\section{Introduction}
The purpose of this paper is to present an introduction to the subject
of the $Z_N$ symmetry\cite{zn} which is present in the study of $SU_N$ Gauge
Theories at finite temperature. I begin with a brief review of the Euclidean
Path Integral formulation of pure $SU_N$ Gauge Theories (without
fermions)  and of the Wilson--Polyakov
Line which is used as an order parameter for this theory. This is
followed by a discussion of the $Z_N$ symmetry, its relationship
to Confinement and the breaking of this symmetry at high temperatures.
$Z_N$ domain walls and $Z_N$ bubbles are then introduced. I then
discuss what happens when fermions are introduced including the
presence of metastable configurations in the Euclidean Partition Function.
The paper concludes with a discussion of the physical interpretation of
these metastable states, and of $Z_N$ domains in general, in
Minkowski space.

\section{Pure $SU_N$ Gauge Theories}

Pure $SU_N$ Gauge Theories at a finite temperature $T=\beta^{-1}$
are usually studied via the Partition Function\cite{ftZ}
\equation
	Z(\beta)~=~{\rm Tr}~{\rm e}^{-\beta H}~
	\propto~\int_{A_i(\tau=0)=A_i(\tau=\beta)}
	{\cal D}A_\mu ~ {\rm exp}\left[-S_E(\beta)\right]
\endequation
with the Eucliean Action $S_E(\beta)$ given by
\equation
	S_E(\beta)~=~{1\over g^2}\int_0^\beta d\tau\int d^3x~
	{1\over 4}~F_{\mu\nu}^aF^{\mu\nu}_a
\endequation
where $F_{\mu\nu}^a=\partial_\mu A_\nu^a-\partial_\nu A_\mu^a
+(A_\mu \times A_\nu)^a$ are the Field Strengths. The above expression
for $Z(\beta)$ is derived by considering only the states $\vert\psi>$ which
satisfy Gauss' Law $D_iE_i\vert\psi>=0$. It thus represents the
Partition Function for the theory in the {\bf absense} of any
external sources. The Free Energy of such a system is given by
$F(\beta)=-(1/\beta){\rm log}\left[Z(\beta)\right]$.

The Partition Function in the presence of a single external ``quark''
source (i.e. a static source in the fundamental representation of
$SU_N$) at a spatial point $\vec x$  is given by\cite{wpline}
\equation
	Z_q(\beta)~
	\propto~\int_{A_i(\tau=0)=A_i(\tau=\beta)}
	{\cal D}A_\mu ~ {\rm exp}\left[-S_E(\beta)\right]
	~\times~ \left[{\rm Tr}~L(\vec x)\right]
\endequation
where
\equation
	L(\vec x)~=~\left[ {1\over N}~P{\rm exp}\left(
	i\int_0^\beta A_0(\vec x,\tau) d\tau\right)\right]
\endequation
is called the Wilson--Polyakov Line.

Under a gauge transformation $U(\vec x, \tau)$, $L(\vec x)$ transforms as
\equation
	L(\vec x)\rightarrow U(\vec x,0)L(\vec x)U(\vec x,\beta)
\endequation
It thus follows that ${\rm Tr}~L(\vec x)$ is invariant under gauge
transformations which are {\bf periodic} in time i.e.
$U(\vec x,0)=U(\vec x,\beta)$. The increase in Free Energy $\delta F$
when adding
such an external source is thus given by
\equation
	{\rm e}^{-\beta\left(\delta F\right)}~=~\langle{\rm Tr}~L(\vec x)
	\rangle~=~
	{{Z_q(\beta)}\over{Z(\beta)}}
\endequation
Similarly the excess Free Energy
$V(\vert\vec x-\vec y\vert)$ of a ``quark'' and an ``antiquark''
source at locations $\vec x$ and $\vec y$ respectively is given by
\equation
	\langle {\rm Tr}~L^\dagger(\vec y)~{\rm Tr}~L(\vec x)\rangle
	\propto {\rm e}^{-\beta~V(\vert\vec x-\vec y\vert)}
\endequation

It follows from the above discussion that $\langle {\rm Tr}~L(\vec x)\rangle$
is a useful order parameter for distinguishing a confining from a deconfining
phase in Gauge Theories without fermions. If $\langle {\rm Tr}~L\rangle=0$
then $\delta F$ is infinite and it costs an infinite amount of energy
to introduce a single source. Furthermore $\langle {\rm Tr}~L^\dagger
(\vec y)~{\rm Tr}~L(\vec x)\rangle\rightarrow 0$ as $\vert\vec x-\vec y\vert
\rightarrow \infty$ so that $V(\vert\vec x-\vec y\vert)\rightarrow \infty$
at large distances. This signals a confining phase. If, on the other hand,
$\langle {\rm Tr}~L\rangle \ne 0$ then the potential energy $V$ is finite
at large distance. This signals a deconfining phase.

\vfill
\eject

\section {$Z_N$ Symmetry}

It so happens that the Euclidean Path Integral possesses a
discrete symmetry which implies that $\langle {\rm Tr}~L\rangle = 0$.
For $SU_N$ this symmetry transforms
\equation
	L\rightarrow {\rm e}^{2\pi i {\rm k}/N} \times L \label{ltrans}
\endequation
for $k=1,2...N$ leaving the action $S_E$ invariant. To see this note that
in our formalism the Partition Function is a sum over only {\bf periodic}
Gauge Potentials $A_i$. If, however, we consider a nonperiodic Gauge
Transformation $U(\vec x,\tau)$ such that
\equation
	U(\vec x,\tau=0)=I;~~~~U(\vec x,\tau=\beta)=M(\vec x)\ne I
	\label{zntrans}
\endequation
then the periodic boundary conditions on $A_i$ will be maintained if
and only if $M(\vec x)$ commutes with {\bf any} $A_i$. This can only happen
if $M$ is in the Center of the group $SU_N$
\equation
	M= {\rm e}^{2\pi i {\rm k}/N} \times I
\endequation
for $k=1,2...N$. This is the group $Z_N$. Under this transformation both
the action and the boundary conditions are invariant but the order
parameter $\langle {\rm Tr}~L\rangle$ is not invariant but transforms as
in Eq. (\ref{ltrans}). This then implies that $\langle {\rm Tr}~L\rangle =0$
which seems to imply confinement. The loophole is that this symmetry is
a discrete, global symmetry and thus it is entirely possible that
this symmetry is spontaneously broken at some temperatures.
It thus follows that the issue of Confinement is intimately tied
to the question of whether the $Z_N$ symmetry is spontaneously broken.
Spontaneous breaking of the symmetry leads to $\langle {\rm Tr}~L\rangle \ne 0$
which signals a deconfined phase.

It has been shown\cite{znbrkg} that the $Z_N$ symmetry is indeed broken
in weak coupling perturbation theory which is expected to be
valid at high temperatures. The basic reason for this is that
perturbation theory is an expansion around $A_\mu =0$ which implies
that $\langle {\rm Tr}~L\rangle=1$-(perturbative corrections).
Perturbative calculations
of the Effective Potential for $A_0$ or, equivalently for $L$ show
a minimum at $\langle {\rm Tr}~L\rangle=1$ and at the $Z_N$ symmetric
points $\langle {\rm Tr}~L\rangle={\rm exp}(2\pi i{\rm k}/N)$.
A typical curve (for $SU_2$) is shown in Figure 1.
The presence of this minimum in the deconfined (high temperature) phase
and its absense in the confined (low temperature) phase is
confirmed by numerical Lattice simulations.

It is interesting to note that, unlike most other symmetries, this $Z_N$
symmetry is broken in the low temperature phase and unbroken in
the high temperature phase.

\section{Domain Walls and Bubbles}

The form of the Effective Potential shown in Figure 1
in the high temperature phase implies the
existence of domain walls in the Euclidean Path Integral. If we imagine
forcing boundary conditions on our system such that for $x_3\rightarrow
-\infty$
the system sits near the minimum at $\langle {\rm Tr}~L\rangle=1$ whereas
for $x_3\rightarrow +\infty$ the system is near another of the minima of
the Effective Potential, say at $\langle {\rm Tr}~L\rangle=
{\rm exp}(2\pi i{\rm k}/N)$
then there will be a Domain Wall separating these two
``vacua''.  Calculations of these Domain Wall energies have been carried
out both in perturbation theory\cite{pka} and numerically and
they will be discussed in another paper at this workshop by C.
Korthals--Altes.

\vspace{1in}
\epsfysize=4in
\epsfbox[-125 0 624 576]{fig1.ps}

\vspace{-1in}
\centerline{\bf Figure 1}
\centerline{\bf Free energy versus $q\propto {\rm log}
	\langle {\rm Tr}~L\rangle$ for $SU(2)$ with no Fermions}

\vskip .4in

A consequence of the existence of these domain walls is that there also
exist $Z_N$ bubbles. These bubbles are regions in {\bf space}
for which $\langle {\rm Tr}~L\rangle \simeq 1$ outside
the bubble but at the center of the bubble
$\langle {\rm Tr}~L\rangle \simeq  {\rm exp}(2\pi i{\rm k}/N)$.
As the temperature $T$ decreases it becomes more likely to form these
bubbles. This happens because the probability to form a bubble
is proportional to ${\rm exp}(-S/g^2)$. The action is proportional
to $T^4$ and $g^2$ increases as $T$ is decreased. As the temperature
is lowered more of these bubbles will be present and they will eventually
``randomize'' $L$ until at some critical temperature the symmetry
is restored and $\langle {\rm Tr}~L\rangle =0$. This is the
confining--deconfining transition.

\section {Gauge Theories with Fermions}

The situation described above for pure Gauge Theories change
significantly when Fermions (in the fundamental representation of
$SU_N$) are introduced. We consider a theory with $N_f$ flavours
of ``quarks'' described by Fermionic fields $\psi$. The Partition
Function is given by
\equation
	Z_f(\beta)~=~{\rm e}^{-\beta F_f(\beta)}~
	\propto~\int_{\buildrel{A_i(0)=A_i(\beta)}\over
	{\psi(0)=-\psi(\beta)}}
	{\cal D}A_\mu {\cal D}\psi {\cal D}\psi^\dagger
	~ {\rm exp}\left[-S_E^f(\beta)\right]
\endequation
where $S_E^f(\beta)$ is the usual Euclidean Action for QCD.
Note the antiperiodic boundary conditions which $\psi$ satisfies.

We expect, both on physical and on mathematical grounds, that
$\langle {\rm Tr}~L\rangle \ne 0$ in this case. Physically this is a
result of the fact that the potential energy $V(r)$ of two static sources
separated by a distance $r$ does not grow as $r\rightarrow\infty$ due
to screening by the dynamical fermions. Furthermore the ground
state in the presence of a single quark source has a finite energy
since it includes a dynamical antiquark to which it is bound.

This physical expectation is realized mathematically by the fact that
there is no $Z_N$ symmetry for this system and thus there is no
mathematical reason to suppose that $\langle {\rm Tr}~L\rangle = 0$.
(There is, in fact, no known order parameter which distinguishes
a confining from a deconfining phase in the presence of quarks.)
The reason for the loss of $Z_N$ symmetry is that even though
both the Action and the periodic boundary conditions on $A_i$ are
maintained by the transformation (\ref{zntrans}) the antiperiodic
boundary conditions on $\psi$ are not preserved since
\equation
	\psi(\beta)\rightarrow U(\beta)\psi(\beta) =
	- {\rm e}^{2\pi i{\rm k}/N}\psi(0)
\endequation
which is a simple reflection of the fact that fields transforming
under the fundamental representation of $SU_N$ are not invariant
under its center.

This lack of $Z_N$ invariance can be vividly demonstrated by
computing the Effective Potential for $\langle {\rm Tr}~L\rangle $.
The specific case of $N=3$ is plotted in Figure 2 for various
values of $N_f$.  In this figure $F_f/T^4$ is plotted as a function
of a variable $q\propto {\rm log}(\langle {\rm Tr}~L\rangle)$.
Note the presence of ``metastable minima'' at $q=1$ and $2$.
These minima actually become maxima for sufficiently large $N_f$
although this is not shown in the figure.

\section {Interpretation of the $Z_N$ Domains in Minkowski Space}

It is tempting to treat the metastable extrema of an Effective Potential
such as that of Figure 2 as a physical metastable state in Minkowski
space. There have been several attempts to do this and interesting
physical and cosmological consequences of this have been
suggested\cite{dolk}. Despite these attempts I believe that
these metastable {\bf Euclidean} extrema have no {\bf direct}
physical interpretation\cite{bksw}.

First note that if we {\bf do} interpret these metastable states
physically we run into very serious trouble. The reason is that
the Free Energy $F_f\propto T^4$. For example for $SU_3$ with
$N_f=4$, the metastable extremum at $q=1$ has a free energy
$F=\gamma T^4$ {\bf with $\gamma > 0$}. This happens for a very
large class of values for $N_f$ and $N$.  This is a disaster if these
points represent true metastable states of the system. The reason is
that if $F=\vert \gamma \vert T^4$ then the pressure
$p=-\vert \gamma \vert T^4$, the internal energy $E=-3\vert \gamma \vert T^4$,
the specific heat $c=-12\vert \gamma \vert T^3$ but worst of all the
entropy
\equation
	S~=~{{E-F}\over T}~=~-4\vert \gamma \vert T^3
\endequation
This negative entropy implies, among other things, that there
is less than one state available to the system. The above
values for the thermodynamic quantities are clearly unphysical.

\epsfysize=4in
\epsfbox[-125 0 624 576]{fig2.ps}

\vskip .2in
\centerline{\bf Figure 2}
\centerline{\bf Free energy versus $q\propto {\rm log}
	\langle {\rm Tr}~L\rangle$ for $SU(3)$ with various numbers}
\centerline{\bf  of fermion flavours}

\vskip .4in

This problem (which is caused by the ``wrong'' sign of $F$) can
{\bf artificially} fixed by adding extra particles to the system
whose entropy $S>0$. This is what happens, for example, in the
Standard Electoweak Model. But this does not solve the fundamental
problem that a physical interpretation of these states is
untennable.

The basic problem is that $\langle {\rm Tr}~L\rangle$ is {\bf fundamentally}
a Euclidean object. It is, in fact, nonlocal in Euclidean time.
In fact a Euclidean $A_0\ne 0$ (which leads to a nontrivial
$\langle {\rm Tr}~L\rangle)$ corresponds to an {\bf imaginary}
$A_0$ in Minkowski space. Thus in Minkowski space a constant $A_0$ looks
like a purely imaginary chemical potential for ``color'' charge!
This is clearly unphysical for a Minkowski object. In fact if we use
as an example the case when $N$ is even and $q=N/2$
(i.e. $\langle {\rm Tr}~L\rangle) = -1$ then the Fermi distribution
in such a constant background $A_0$ field turns into a Bose
distribution but with
\equation
	n(p)~=~ -{1\over{{\rm e}^{\beta E(p)}-1}}
\endequation
Note the minus sign in front of the expression! This is related to the
negative entropy of the system.

The lesson of the above discussion is that $Z_N$ bubbles should
{\bf not} be interpreted as physical bubbles any more than
Instantons should be interpreted as real Minkowski objects.
But these $Z_N$ bubbles {\bf do} contribute to the Partition Function
and to expectation values of observables
if they are calculated using the Euclidean Path Integral. They should
thus be included in a non--perturbative analysis of the thermodynamics
of QCD. The physical interpretation of these bubbles is further
discussed in a paper by A.V. Smilga presented at this Workshop.

\section{Summary}

\noindent
-- The $Z_N$ symmetry for $SU_N$ gauge theories plays an important role
in the {\bf Euclidean} analysis of its thermodynamics.

\medskip
\noindent
-- $Z_N$ bubbles
are likely to play a role in the confining--deconfining phase
transition.

\medskip
\noindent
-- Fermions in the Fundamental representation break the
$Z_N$ symmetry.

\medskip
\noindent
-- The resulting metastable extrema of the Effective Potential
should {\bf not} be interpreted as physically attainable states
nor should the $Z_N$ bubbles which attain these extrema in their
cores.

\medskip
\noindent
-- These metastable extrema {\bf do} however contribute to the
thermodynamics of the system.


\vspace{1cm}

\noindent
{\bf REFERENCES}

\begin{thebibliography}{9}

\bibitem{zn} G. 't Hooft, Nucl. Phys. {\bf B138} 1 (1987)

\bibitem{ftZ} See for example D.J. Gross, R.D. Pisarski, and
	L.G. Yaffee, Rev. Mod. Phys. {\bf 53} 43 (1981) and
	references therein.

\bibitem{wpline} A.M. Polyakov, Phys. Lett. {\bf 72B}  477 (1978);
		B.Svetitsky and L.G. Yaffe, Nucl. Phys. {\bf B210} 423 (1982);
	L. McLarren, B. Svetitsky, Phys. Rev. {\bf D24} 450 (1981);
	J. Kuti, J. Polonyi, K. Szlachanyi, Phys. Lett. {\bf 98B}
	199 (1981)


\bibitem{znbrkg} Nathan Weiss,Phys. Rev. {\bf D24} 475 (1981);
	Phys. Rev. {\bf D25} 2667 (1982)


\bibitem{pka} T. Bhattacharya, A. Gocksch, C. Korthals Altes
	and R.D. Pisarski, Phys. Rev. Lett. {\bf 66} 998 (1991);
	Nucl. Phys. {\bf B383} 497 (1992)


\bibitem{dolk} V. Dixit, M.C. Ogilvie, Phys. Lett. {\bf B269}
	353 (1991); J. Ignatius, K. Kajantie, K. Rummakainen,
	Phys. Rev. Lett. {\bf 68} 737 (1992)

\bibitem{bksw} V.M. Belyaev, I. Kogan, G.W. Semenoff, Nathan Weiss,
	Phys. Lett. {\bf B277} 331 (1992)

\end{thebibliography}
\end{document}